\documentclass[aps,prfluid,groupedaddress,reprint,citeautoscript]{revtex4-1}

\usepackage{graphicx}
\usepackage{amsfonts}
\usepackage{amsmath}
\usepackage{amssymb}
\usepackage{bm}                    % boldmath
\usepackage{dcolumn}               % align table by decimal
\usepackage[multiple]{footmisc}    % comma between consecutive footnotes
\usepackage{setspace}              % boldmath
\usepackage{cancel}                % cancel
\allowdisplaybreaks[3]  %  splits long equation over two pages
\usepackage{color}
\newcommand{\Eq}    [1] {Eq.~(\ref{#1})}       % mid sentence
\newcommand{\Eqs}   [1] {Eqs.~(\ref{#1})}
\newcommand{\eq}    [1] {(\ref{#1})}
\newcommand{\Sec}   [1] {Section~\ref{#1}}     % mid sentence

\newcommand{\eg}        {\textit{e.g.}}

\newcommand{\ie}        {\textit{i.e.}}
\newcommand{\supscr} [1]{\textsuperscript{#1}}

\newcommand{\lhs}       {l.h.s.}
\newcommand{\rhs}       {r.h.s.}
\newcommand{\supth}     {\supscr{th}}          % usage: $i${\supth}

\newcommand{\posit}      {x}  % one  position
\newcommand{\equi}       {\mathrm{ eq}}

\newcommand{\veloc}      {U}
\newcommand{\angvel}     {\varOmega}

\newcommand{\flov}       {u}
\newcommand{\flow}       {\omega}
\newcommand{\posFl}      {r}
\newcommand{\stressFl}   {s} % \sigma}

\newcommand{\forc}       {F}

\newcommand{\stress}     {S}
\newcommand{\sTress}     {s}
\newcommand{\strain}     {E}
\newcommand{\mblty}      {M} % {\mu}
\newcommand{\rstnc}      {R} % {\xi}

\newcommand{\Epot}       {\Phi}

\newcommand{\Brown}      {B} % \mathrm{B}}
\newcommand{\BrowN}      {\mathcal{\Brown}} % \mathrm{b}}
\newcommand{\hydro}      {H} % \mathrm{H}}
\newcommand{\moblty}     {M} % \mathrm{M}}
\newcommand{\thermo}     {T} % \mathrm{T}}
\newcommand{\Dirac}      {\delta}
\newcommand{\Kron}   [1] {\delta_{#1}}

\newcommand{\bcddot}     {\mathbf{:}}
\newcommand{\cross}      {\times}
\newcommand{\dyad}       {\otimes}

\newcommand{\diffcoef}   {D}
\newcommand{\fric}       {\gamma}
\newcommand{\grav}       {g}
\newcommand{\velGrav}    {v_\grav}
\newcommand{\kB}         {k_B}
\newcommand{\rkBT}       {\beta}
\newcommand{\radCol}     {a}
\newcommand{\visc}       {\eta}
\newcommand{\volBox}     {V}
\newcommand{\Zero}       {\boldsymbol{0}}
\newcommand{\ZERO}       {\mathbf    {0}}

\newcommand{\ONE}        {\mathbf    {1}}

\newcommand{\Trnspsd}    {\mathrm{T}}
\newcommand{\XXX}        {\mathbf{X}}
\newcommand{\Dt}         {\Delta t}
\newcommand{\DtSq}       {\sqrt{\Dt}}

\newcommand{\ProbDistF}  {P}

\newcommand{\zO}         {z_0}

\newcommand{\mass}       {m}

\newcommand{\Ncol}       {N}
\newcommand{\Proj}       {\hat{\mathcal{P}}}

\newcommand{\rand}       {\theta}

\newcommand{\Pos}        {\boldsymbol{\posit}}
\newcommand{\Posi}    [1]{\Pos_{#1}}
\newcommand{\Posdt}      {\dot{\Pos}}

\newcommand{\DPosBrwn}   {\Delta \Pos^\Brown}

\newcommand{\diff}       {\diffcoef}
\newcommand{\Diff}       {\mathbf{\diffcoef}}

\newcommand{\Vel}        {\boldsymbol{\veloc}}
\newcommand{\Veli}    [1]{\Vel_{#1}}

\newcommand{\VelAv}      {\bar{\Vel}}
\newcommand{\Av}         {\boldsymbol{\angvel}}  %  how to make italic?

\newcommand{\viscFld}    {\visc_0}
\newcommand{\viscInf}    {\visc_\infty}

\newcommand{\Flv}        {\boldsymbol{\flov}}
\newcommand{\Flw}        {\boldsymbol{\flow}}

\newcommand{\FlvO}       {        \Flv _0  }
\newcommand{\PosFl}      {\boldsymbol{\posFl}    }

\newcommand{\Stn}        {\mathbf{\strain}    }
\newcommand{\StsFl}      {\mathbf{\stressFl}}

\newcommand{\Frc}        {\boldsymbol{\forc}}
\newcommand{\FEf}        {\tilde     {\Frc}}
\newcommand{\FAv}        {\bar       {\Frc }}

\newcommand{\FrcBrwn}    {\Frc  ^\Brown}
\newcommand{\FrcBrAv}    {\FAv{}^\Brown}
\newcommand{\FrcHyd}     {\Frc  ^\hydro}
\newcommand{\FrcMbl}     {\FEf{}^\moblty}
\newcommand{\FrcPot}     {\Frc  ^\Epot}
\newcommand{\FrcThrm}    {\FEf{}^\thermo}
\newcommand{\Frci}    [1]{\Frc          _{#1}}
\newcommand{\FrciBrwn}[1]{\Frc  ^\Brown _{#1}}

\newcommand{\FrciMbl} [1]{\FEf{}^\moblty_{#1}}
\newcommand{\FrciPot} [1]{\Frc  ^\Epot  _{#1}}
\newcommand{\FrciThrm}[1]{\FEf{}^\thermo_{#1}}

\newcommand{\Sts}        {\mathbf{\stress}}
\newcommand{\STs}        {\mathbf{\sTress}}
\newcommand{\SEf}        {\tilde   {\Sts}}
\newcommand{\StsAv}      {\bar     {\Sts}}
\newcommand{\Stsi}    [1]{\Sts          _{#1}}
\newcommand{\StsiBrwn}[1]{\Sts  ^\Brown _{#1}}
\newcommand{\StsBrFr}    {\Sts  ^\BrowN}
\newcommand{\StsBrwn}    {\Sts  ^\Brown}
\newcommand{\StsHyd}     {\Sts  ^\hydro}

\newcommand{\StsMbl}     {\SEf{}^\moblty}
\newcommand{\StsPot}     {\Sts  ^\Epot}
\newcommand{\StsStn}     {\Sts  ^\strain}
\newcommand{\StsThrm}    {\SEf{}^\thermo}

\newcommand{\StsiStn} [1]{\Sts  ^\strain_{#1}}
\newcommand{\StsTot}     {\STs  ^\Sigma}
\newcommand{\pHyd}       {p}

\newcommand{\Mblty}      {\mathbf{\mblty}}
\newcommand{\Mblvf}      {\Mblty_{\veloc  \forc  }}
\newcommand{\Mblvn}      {\Mblty_{\veloc  \strain}}
\newcommand{\Mblsf}      {\Mblty_{\stress \forc  }}
\newcommand{\Mblsn}      {\Mblty_{\stress \strain}}
\newcommand{\MblvfSq}    {\Mblvf^{ 1/2}}
\newcommand{\MblvfI}     {\Mblvf^{-1}}

\newcommand{\CCi}        {\mathbf{C}_i}

\newcommand{\Rstnc}      {\mathbf{\rstnc}}
\newcommand{\Rstfv}      {\Rstnc_{\forc   \veloc }}
\newcommand{\Rstfn}      {\Rstnc_{\forc   \strain}}
\newcommand{\Rstsv}      {\Rstnc_{\stress \veloc }}
\newcommand{\Rstsn}      {\Rstnc_{\stress \strain}}
\newcommand{\RstfvI}     {\Rstfv^{-1}}

\newcommand{\pdf}        {\ProbDistF}
\newcommand{\pdfEq}      {\pdf_{\equi}}
\newcommand{\pdfO}       {\pdf_0}
\newcommand{\pdfGreen}   {\pdf} % \mathcal{\pdf}}
\newcommand{\chempot}    {\mu}

\newcommand{\flux}       {J}
\newcommand{\Flux}       {\boldsymbol{\flux}}

\newcommand{\Rndi}    [1]{\boldsymbol{\rand}_{#1}}

\newcommand{\dfdx}    [2]{\frac{ \partial #1   }{ \partial #2   }}
\newcommand{\ddfdxx}  [2]{\frac{ \partial^2 #1 }{ \partial #2^2 }}

\begin{document}

\title{The contribution of Brownian motion
       to the stress in a colloidal suspension}

\author{Duraivelan Palanisamy}
\author{Wouter K. den Otter}
  {\email{w.k.denotter@utwente.nl}
        }
\affiliation{
  Multi Scale Mechanics,
  Faculty of Engineering Technology and MESA+ Institute for Nanotechnology,
  University of Twente,
  Enschede,
  The Netherlands.
}

\date{\today}

\begin{abstract}
The deviatoric stresses of colloidal suspensions are routinely calculated,
   in theoretical studies
      as well as in Brownian and Stokesian Dynamics simulations,
  using the expression introduced by
    Batchelor [\textit{J. Fluid Mech.} \textbf{83}, 97--117 (1977)].
We show by example that the central feature of its derivation,
    the thermodynamic force
      representing the mean Brownian motion of colloids
        as flow against the density gradient,
  is inconsistent with the motion of the colloids
    on the Smoluchowski time scale.
The mean Brownian motion is well-known to originate
    in the spatial variation of the grand mobility matrix,
% as implemented in Brownian and Stokesian Dynamics simulations,
  which therefore ought to be included
    in stress calculations instead.
A novel expression for the stress is derived,
 restoring the hydrodynamic relation between
   the motion of suspended particles and their induced stress.
\end{abstract}

\maketitle

%======================================================================

\section{Introduction\label{sec:intro}}

Adding rigid colloids or flexible polymers to simple fluids
    is well-known to affect the flow behaviour of the fluid,
  raising the viscosity and giving rise to visco-elastic phenomena
    like shear thinning;
  the interested reader is refered to various reviews
      \cite{Bird2,DoiEdwards,KimKarrila,Larson,GuazzelliMorris}.
Einstein \cite{Einstein05,Einstein06} famously derived that
  the viscosity of a dilute suspension of spherical colloids
    increases linearly with the colloidal volume fraction.
Batchelor \cite{Batchelor76,Batchelor77} proposed
%   in a winding and speculative manner,
  a general expression for the deviatoric stress
    of non-dilute colloidal suspensions of spherical colloids
      subject to Brownian motion,
  which has become widely accepted as the standard expression
    in theoretical and simulation studies.
It is therefore disconcerting to note that
  the forces entering Batchelor's stress calculation
  differ from those entering the equations of motion of the colloids
    \cite{Ermak78,DoiEdwards,Kampen,Ottinger}.
The problems comprise % are tracked down to
  the use of a `thermodynamic force' acting on the particles
  and the omission of a subtle correction
    for the configuration-dependence of the hydrodynamic matrix.
A new expression for the deviatoric stress is derived
 by combining a number of well-known results
   on micro-hydrodynamics and Brownian motion.

Batchelor \cite{Batchelor77} derived
  the stress of a colloidal suspension of Brownian particles as
\begin{subequations}
\begin{align}
  \label{eq:Batchelor1}
    \StsTot
 &=
  -
    \pHyd
    \ONE
  +
    2
    \viscFld
    \Stn
  +
    \frac{ 1 }{ \volBox }
    \sum_i % {i=1}^{\Ncol}
      \StsiStn{i}
  -
    \frac{ 1 }{ \volBox }
    \sum_i % {i=1}^{\Ncol}
      \left(
        \CCi
        +
        \Posi{i}
        \ONE
      \right)
      \cdot
      \Frci{i}
%     \left(
%       \FrciPot {i}
%     +
%       \FrciThrm{i}
%     \right)
\\
  \label{eq:Batchelor2}
 &=
  -
    \pHyd
    \ONE
  +
    2
    \viscFld
    \Stn
  +
    \frac{ 1 }{ \volBox }
    \sum_i % {i=1}^{\Ncol}
      \StsiStn{i}
  -
    \frac{ \kB T }{ \volBox }
    \sum_i % {i=1}^{\Ncol}
      \nabla_i
      \cdot
      \CCi,
% -
%   \sum_i
%     \Posi{i}
%     \dyad
%     \FrciPot{i}
% -
%   \sum_i
%     \CCi
%     \cdot
%     \FrciPot{i},
\end{align}
\end{subequations}
  where, in the first line,
    $\pHyd$ is the hydrostatic pressure,
    $\viscFld$ denotes the viscosity of the suspending fluid,
    $\Stn$ is the imposed uniform strain rate tensor,
    $\volBox$ is the volume,
    the summations run over all particles, %  or particle pairs,
    $\StsiStn{i}$ is the hydrodynamic stress on colloid $i$
      due to the strain deformation
        (in the absence of Brownian contributions),
    the product $\Stn \bcddot \CCi$ yields
      the velocity boost of said particle in a strain deformation
        due to hydrodynamic interactions with the surrounding colloids,
%   $\kB$ is Boltzmann's constant,
%   $T$ stands for the absolute temparuture,
    $\Posi{i}$ is the position of the $i$\supscr{th} particle,
%   $\FrciPot{i}$ is a conservative force acing on the particle
%     (not included by Batchelor),
%   and, finally,
%   $\FrciThrm{i}$ is the thermodynamic force on that particle.
    and
    $\Frci{i}$ denotes the non-hydrodynamic force acting on the particle.
Batchelor equates % \textit{assumed}
  this force to the `thermodynamic force', $\FrciThrm{i}$,
    an effective (denoted by the tilde) force describing
      the average result of Brownian motion
        as a deterministic motion against the concentration gradient;
  a more extensive discussion of this force is presented below.
He thus arrived at the second line in the above equation,
    where $\kB$ is Boltzmann's constant
      and $T$ stands for the absolute temperature,
  as an approximate expression to the stress
    valid at low volume fraction only.
Felderhof and Jones \cite{Felderhof87a}
    and N\"agele and Bergenholtz \cite{Nagele98}
  added a conservative potential-based force
      $\FrciPot{i} = - \nabla_i \Epot$,
    describing direct interactions between the colloids,
      $\Frci{i} = \FrciThrm{i} + \FrciPot{i}$,
  thereby arriving at a stress expression
    with an inter-colloidal virial contribution,
\begin{align}
  \begin{split}
  \label{eq:Nagele}
    \StsTot
 &=
  -
    \pHyd
    \ONE
  +
    2
    \viscInf
    \Stn
% +
%   \frac{ 1 }{ \volBox }
%   \sum_i % {i=1}^{\Ncol}
%     \StsiStn{i}
  -
    \frac{ \kB T }{ \volBox }
    \sum_i % {i=1}^{\Ncol}
      \nabla_i
      \cdot
      \CCi
\\
 &\;\;\;
  -
    \sum_i
      \Posi{i}
      \dyad
      \FrciPot{i}
  -
    \sum_i
      \CCi
      \cdot
      \FrciPot{i},
  \end{split}
\end{align}
  where $\viscInf$ is the high frequency limiting viscosity;
  this expression was shown to hold true at all volume fractions.
These equations and equivalent formulations are widespread
    in the literature,
  both in theoretical and numerical studies.
Our objectives here are
  to discus a problem underlying
    the inclusion of the average Brownian force
      in \Eqs{eq:Batchelor2} and \eq{eq:Nagele},
  and to provide a corrected expression for the stress.

This paper is structured as follows.
Batchelor's argument for the thermodynamic force is repeated
  in \Sec{sec:background}.
In \Sec{sec:revisited} it is shown that
  the thermodynamic force does not agree with
    the motion of colloids on the Smoluchowski time scale,
  and therefore ought not to be included
    in the derivation of a stres expression.
There is a subtle Brownian contribution
  to the motion of colloids on the Smoluchowski time scale
    that must be included instead,
  as explained and used in \Sec{sec:stress}
    to derive a novel stress expression.
The new stress expression is compared against current expressions
    in \Sec{sec:comparison},
  followed by a summary of our findings in \Sec{sec:DiscConc}.
We apologize % in advance
  to readers familiar with more stringent derivations
    of the various partial results
      used to refute Batchelor's argument
        and to derive a new stress expression:
%     to be combined here into the final stress expression:
  we use simple arguments to highlight the mayor points
    that have been systematically overlooked for the last 40 years.

%======================================================================

\section{Background\label{sec:background}}

In a famous \textit{Gedankenexperiment},
  Einstein \cite{Einstein05} considered
    the equilibrium probability distribution function $\pdf(\Pos)$
      of a dilute suspension of identical colloidal particles
        in an external potential $\Epot(\Pos)$.
% It is well-known that the equilibrium at constant temperature $T$
%   is described by Boltzmann's distribution,
% \begin{equation}
%   \label{eq:Boltzmann}
%     \pdf(\Pos)
%   =
%     \pdfO
%     e^{ - \rkBT \Epot(\Pos) },
% \end{equation}
%   where $\rkBT = 1 / ( \kB T )$ with Boltzmann constant $\kB$.
In equilibrium,
  the particle flux due to the external potential
%     $\FrcPot = - \nabla \Epot$,
    is balanced by the flux due to
      Fickian diffusion against the concentration gradient.
The macroscopic flux $\Flux(\Pos)$ then vanishes at every point $\Pos$,
  following
\begin{equation}
  \label{eq:fluxEinstein}
    \Flux
  =
  - \pdf % (\Pos)
    \Mblty
%   \cdot
    \nabla \Epot
  -
    \Diff
%   \cdot
    \nabla \pdf
  =
    \Zero,
\end{equation}
  where $\Mblty$ denotes the mobility matrix
    and $\Diff$ the diffusion matrix.
By inserting Boltzmann's equilibrium distribution,
\begin{equation}
  \label{eq:Boltzmann}
    \pdf(\Pos)
  =
    \pdfO
    e^{ - \rkBT \Epot(\Pos) },
\end{equation}
    where $\pdfO$ normalizes the distribution
      and $\rkBT = 1 / ( \kB T )$,
%       with Boltzmann constant $\kB$ and absolute temperature $T$,
  Einstein showed that equilibrium implies $\Mblty = \rkBT \Diff$.
Using Stokes's expression
    for the mobility of a spherical particle of radius $\radCol$
      in a fluid of viscosity $\visc$,
  Einstein obtained the diffusion coefficient
\begin{equation}
    \Diff
  =
    \frac{ \kB T }{ 6 \pi \visc \radCol }
    \ONE.
\end{equation}
While the Stokes-Einstein expression
  was derived for colloids subject to an external force,
    it is equally valid for unforced colloids.

In a sequel paper,
  Einstein \cite{Einstein06} likened the action of the diffusive term
    to a force acting on every particle,
  while the central equation he solved was still a flux balance.
Batchelor \cite{Batchelor76} described this interpretation as follows:
  `the particle flux due to Brownian migration is the same here
    as if a certain steady force acted on the particles
    (this force being equal and opposite to
      the external force $\FrcPot = - \nabla \Epot$ that,
          in the equilibrium situation,
        produces a convective flux which balances the diffusive flux).'
% that is, in view of \Eq{eq:Boltzmann},
%   the same as if a steady force
The flux balance in \Eq{eq:fluxEinstein} is rewritten as
%begin{subequations}
\begin{align}
  \label{eq:fluxBatchelorAll}
%   \Flux
%&=
  - \pdf % (\Pos)
    \left(
      \Mblty
%     \cdot
      \nabla \Epot
    +
      \frac{ 1 }{ \pdf }
      \Diff
%     \cdot
      \nabla \pdf
    \right)
%\
% \label{eq:fluxBatchelorA}
 &=
    \pdf % (\Pos)
    \Mblty
%   \cdot
    \left(
      \FrcPot
    +
%     \Diff
      \FrcThrm
    \right)
  =
    \Zero,
\end{align}
%end{subequations}
  which,
      following Batchelor \cite{Batchelor76},
  is `the same as if a steady force
\begin{align}
  \label{eq:ThermoForceAverage}
    \FrcThrm( \Pos )
  =
  - \kB T
    \nabla \ln \pdf( \Pos )
\end{align}
  acted on the particle[s].
It is of course not to be supposed
  that the interaction of a particle
      with the molecules of the surrounding medium
    is literally equivalent to
      the exertion of a steady force on the particle.
When the probability density of the particle position is non-uniform,
  the mean Brownian velocity of a particle,
      conditional upon it being near a point $\Pos$,
    is non-zero simply as a consequence of the fact that
      the particle is more likely to have come
        from a direction in which the probability density increases
        than from one in which it decreases;
  and it is this bias in the statistics of particle velocities at $\Pos$
    (which is quite consistent with zero mean of the Brownian velocity
      of a given particle in the absence of an applied force)
    that is equivalent,
        so far as its effect on the diffusive flux is concerned,
      to the action of the steady force \Eq{eq:ThermoForceAverage}
        on the particle.'
The thermodynamic force experienced by the particles near $\Pos$
%   see \Eq{eq:ThermoForceAverage},
  is said to result in a thermodynamic force on colloid $i$
    given by \citep{Batchelor76,Batchelor77}
\begin{equation}
  \label{eq:ThermoForceIndividual}
    \FrciThrm{i}
  =
  - \kB T
    \dfdx{ }{ \Posi{i} }
      \ln
        \pdf( \Posi{1}, \ldots, \Posi{\Ncol} ),
%       \pdf( \PosN ),
\end{equation}
  where $\pdf( \Posi{1}, \ldots, \Posi{\Ncol} )$
% where $\pdf( \PosN )$
%     with $\PosN = ( \Posi{1}, \ldots, \Posi{\Ncol} )$
    is the joint probability distribution function
      of all $\Ncol$ particles.
%   and $\PosN$ is short hand for $ ( \Posi{1}, \ldots, \Posi{\Ncol} )$
Batchelor \cite{Batchelor77} writes that
  `these thermodynamic forces reproduce the statistical bias
      in the random walks of the particles
    which results from the non-uniformity of the
      joint-probability distribution function.'
This use of the thermodynamic force is widespread in text books
  \cite{Bird2,DoiEdwards,Ottinger,Dhont,GuazzelliMorris}.
Building on the thermodynamic force,
  Batchelor \cite{Batchelor77} derived his widely used expression
    for the deviatoric stress
      in a suspension of spherical Brownian colloids at low strain rate,
        see \Eq{eq:Batchelor2},
  as well as an expression for the viscosity of these suspensions
    up to second order in the colloidal volume fraction.

%======================================================================

\section{The thermodynamic force revisited\label{sec:revisited}}

In the reinterpretation
  of the flux balance % over many particles
    as a force balance, %  on every particle,
  it follows from \Eq{eq:fluxBatchelorAll}
    that the particles in an equilibrium suspension
      experience a vanishing nett force
%       and consequently stay at a constant position.
%    are performing erratic motions around their average positions.
  and consequently hover around an equilibrium position.
But the colloids are obviously not stationary,
  as they are continuously subjected to rapidly fluctuating interactions
    with the surrounding solvent molecules
      in perpetual thermal motion.
These fluctuating Brownian forces
    are not contained in the thermodynamic force,
% as both are functions of the colloidal positions only.
  which is devoid of information on
    the dynamical properties of the solvent,
        like the viscosity,
      or those of the solvent molecules.
%       like their positions and velocities.
Consequently,
  the individual colloids must also be experiencing
    fluctuating Brownian forces $\FrciBrwn{i}$;
  these forces must have
    a vanishing average $\langle \FrciBrwn{i}\rangle = \Zero$
      so as not to alter the force balance of \Eq{eq:fluxBatchelorAll}.
As explained in the above citation of Batchelor,
  these Brownain forces are
    the origin of the diffusive flux in \Eq{eq:fluxEinstein}
      and the thermodynamic force in \Eq{eq:fluxBatchelorAll}.
Randomly fluctuating Brownian forces
  are central to any study on the dynamics of colloids in fluids.
The thermodynamic force, however, is not:
  the dynamics of colloidal particles is routinely solved
    without reference to the thermodynamic force
      \cite{Ermak78,DoiEdwards,Brady88,Kampen,Ottinger}.
% Brady88 does mention including ln P in the contuum eq of motion,
%   as well as Euler being correct up to dt^2.
One obvious complication is that the probability distribution $\pdf$,
    and hence $\FrciThrm{i}$,
  is not at hand in particle-based simulations,
    which require explicit expressions of the forces
      in terms of the particle positions
  --~the work-around is integration by parts over configuration space
    to obtain a regular ensemble average,
      a was used in deriving \Eqs{eq:Batchelor2} and \eq{eq:Nagele}.
But that is not the only cause of concern.

Consider a dilute suspension of identical spherical colloids,
  each with the same positive excess mass $\mass$
    relative to the volume of fluid they displace,
  in a gravity field acting along the negative $z$ direction.
Using the flux $\flux$ defined by the {\lhs} of \Eq{eq:fluxEinstein},
 the evolution of
     the one-dimensional overall probability distribution $\pdf( z, t)$
   follows from the conservation expression
     known as the Fokker-Planck-Smoluchowski equation
       \cite{McQuarrie,Risken,Kampen,Gardiner},  % Risken
\begin{align}
  \label{eq:SmoluGrav}
    \dfdx{ \pdf }{ t }
  =
  - \dfdx{ \flux }{ z }
% - \dfdx{ }{ z }
%   \left(
%   - \frac{ \mass \grav }{ \fric }
%     \pdf % (\Pos)
%   -
%     \diff
%     \dfdx{ \pdf }{ z }
%   \right)
  =
    \frac{ \mass \grav }{ \fric }
    \dfdx{ \pdf }{ z }
  +
    \diff
    \ddfdxx{ \pdf }{ z },
\end{align}
  where in the second step
     the effective mass $\mass$,
     the friction coefficient $\fric$,
     the accelaration by gravity $\grav$
     and the diffusion coefficient $\diff$ are assumed constant.
In dilute systems,
  this equation applies to both the macroscopic concentration profile
  and the probability distribution function of an individual particle.
The equilibrium solution,
    in the presence of a wall restricting the motion to $ z \ge 0$,
  recovers the Boltzmann distribution,
\begin{align}
  \label{eq:BoltzmannGrav}
    \pdfEq(z)
  =
    \pdfEq(0)
    e^{ - \rkBT \mass \grav z },
\end{align}
  as is readily confirmed by using Einstein's relation
    $\diff = \kB T / \fric$.
We focus now on the subset of those particles
  that are at a specific height $\zO$, with $\zO \gg 0$,
    at time 0.
Their probability distribution function at a later time $t$
  is obtained as the Green's function to the Smoluchowski equation
% yielding % the conditional probability distribution
    \cite{Kampen,DoiEdwards,Ottinger},
\begin{align}
  \label{eq:Green}
    \pdfGreen( z, t | \zO )
  =
    \frac{ 1 }{ \sqrt{ 4 \pi \diff t } }
    \exp\left[
      \frac{
        \left[ z - \left( \zO + \velGrav t \right) \right]^2
      }{
        4 \diff t^2
      }
    \right],
\end{align}
  with drift velocity $\velGrav = - \mass \grav / \fric$.
% The colloids have a propensity to sink to the bottom of the container.
This solution expresses that
  the colloids in an equilibrium distribution
%   --~including those in an equilibrium distribution~--
  are not hovering around a constant height
%   as implied by the force balance in \Eq{eq:fluxBatchelorAll},
  but have a propensity to gradually sink to the bottom of the container,
    where the Boltzmann probability distribution reaches its maximum.
Because this subset of colloids is part of a larger system in equilibrium,
  it follows that
    the flux balance of the overall density at the macroscopic level,
%     as the subset is part of a system
%       obeying the equilibrium of \Eq{eq:fluxEinstein},
      as in \Eq{eq:fluxEinstein},
  does not translate into
    a balance of potential and thermodynamic forces
      at the microscopic level,
  as implied in the re-interpretation
    of the flux balance as a force balance in \Eq{eq:fluxBatchelorAll}.
% The equilibrium macroscopic density distribution
%   is not maintained by particles remaining at (nearly) fixed positions,
%     as in a force balance,
%   but by a cancellation of the fluxes
%     induced by the potential and concentration gradient.
Colloids are not subject to
    an effective force that drives them against the concentration gradient;
  instead,
    a concentration gradient
      turns the average effect
          of the random Brownian motion of many particles
        into a macroscopic flux against the concentration gradient.

The inevitable conclusion is that
  colloids do not experience
    the `thermodynamic force' envisaged by Einstein and Batchelor,
  and expressions derived using the thermodynamic force
    are to be considered with care.
Of particular interest here is the stress of a colloidal suspension.
\textit{
Batchelor \cite{Batchelor77},
  following his interpretation of the thermodynamic force
    as the average Brownian force on a particle,
  substituted the non-hydrodynamic force $\Frci{i}$ in \Eq{eq:Batchelor1}
    with the thermodynamic force $\FrciThrm{i}$
},
      see his Eq.~(3.8),
  and after some mathematical steps % to eliminate $\pdf$
    arrived at \Eq{eq:Batchelor2}
      as the stress expression valid for low volume fractions.
% Since $\FrciThrm{i}$ does
Likewise,
  Doi and Edwards \cite{DoiEdwards} emphasize in their Eq.~(3.135) that
    the force entering their virial expression
      `must include[s] the thermodynamic force.'
% \citet{GuazzelliMorris}[p. 170] write that
%   `since the thermodynamic forces in a homogeneous suspension
%       are internalyly generated and sum to zero,
%     their effect again arises through a moment of the force distribution,
%       {\ie} as a stress.'
This step
  % of equating the mean Brownian force with the thermodynamic force
    is made by various authors~--
  see for instance
%   \citet{Russelxx}[Eq.~(xxx)],
    Felderhof \cite{Felderhof87b},
      Eqs.~(4.8), (5.2), (5.6), (5.7) and (7.12),
%   \citet{KimKarrila}[p127-8],
    Wagner \cite{Wagner93}, Eq.~(6),
    Brady \cite{Brady93}, Eq.~(42),
    Strating \cite{Strating95}, Eq.~(A2), and
%   \citet{Dhont}[Eq.~(4.99)] and
    N\"agele and Bergenholtz, \cite{Nagele98}, Eq.~(14)~--
  in deriving stress expressions for colloidal suspensions.
But, as observed above,
\textit{
  the thermodynamic force is unrelated to the motion of the colloids
    and should therefore not be used in deriving a stress expression
}
  --~not in the stress tensor of a given configuration,
    nor in an ensemble average. % over all configurations.
This is not to say that
  non-zero mean Brownian displacements do not contribute to the stress
    --~Batchelor was correct to notice that they are relevant.
But
\textit{the non-zero mean Browninan displacements
  that appear in the Langevin equation of motion,
    and should be used in the stress expression,
    result from the spatial non-homogeneity of the diffusion matrix%
},
      rather than from the thermodynamic force.
A revised expression for the stress % in a colloidal suspension,
%   replacing the thermodynamic force
%     with the correct mean average Brownian force,
  will be derived in the next section.

For completeness,
    the colloids initially at $\zO$ will evidently not sink forever.
Interactions with the wall % at $z = 0$
    will eventually cause deviations from \Eq{eq:Green},
  and for long times the conditional probability
    converges to the equilibrium distribution,
      $\pdfGreen( z, \infty | \zO ) = \pdfEq( z )$.
This limiting behaviour is only obtained on a very long time scale,
  with the particles bouncing off the wall numerous times.
For a macroscopic system in equilibrium,
    comprising many particles interacting with the wall at any moment,
  both the flux balance of \Eq{eq:fluxEinstein}
    and a force balance between gravity and wall forces
  are obeyed nearly instantaneously.
For an individual particle in this system, however,
  its force balance between gravity and wall forces
%     required to make the particle hover
    is reached only on a very long time scale;
  on the far shorter Smoluchowski time scale
      of motion over a fraction of the colloid's size,
    the persistent pull by gravity results in
      a downward mean motion (that is, along the concentration gradient)
        with super-imposed Brownian fluctuations.
In summary,
  the thermodynamic force based on the gradient of $\kB \ln \pdf(\Pos)$,
    or on the gradient of $\chempot(\Pos) / T(\Pos)$
      with $\chempot$ the chemical potential,
    provides an effective force in the phenomenological relation
      for the evolution of the macroscopic concentration profile
        \cite{GrootMazur},
  but this force is not to be applied to individual colloids.

%======================================================================

\section{The stress\label{sec:stress}}

We now set forth to derive an expression for
  the deviatoric stress in a suspension of rigid Brownian colloids
    in the Stokesian limit,
      {\ie} for vanishing Reynolds and Stokes numbers,
  by combining a couple of well-known results
    on micro-hydrodynamics and Brownian motion.
% For didactical purposes,
To keep the exposition % relatively straightforward,
    focussed on the key issues,
% In order not to overburden the derivation with irrelevant details,
% To simplify the explanation,
  and to facilitate the comparison
%   with the earlier work by \citet{Batchelor77},
    with earlier work
      \cite{Batchelor77,Felderhof87a,DoiEdwards,Nagele98},
  we initially restrict the discussion to linear velocities only
% --~the inclusion of angular velocities in the final result
%     will be explained towards the end of this section.
  --~the inclusion of angular velocities
      will be postponed till \Sec{AngVel}.
We will start by repeating a couple of well-known results,
  to describe the background and set the notation,
  before merging them into an expression for the stress.

%----------------------------------------------------------------------

\subsection{Stokesian flow}

Consider % first
  an isolated non-Brownian particle in a Newtonian fluid.
The particle is described by
    its position $\Pos$ and velocity $\Vel = \Posdt$;
  the externally imposed macroscopic linear flow field is given by
      $\Flv(\PosFl) = \FlvO + \Stn \PosFl$ % at position $\PosFl$
    with a constant small strain rate $\Stn$,
      {\ie} the traceless symmetric $( 3 \times 3 )$
        velocity gradient matrix.
It is well-known from micro-hydrodynamics that
  the hydrodynamic drag force $\FrcHyd$ and stress $\StsHyd$
      experienced by the colloid
    are related under Stokesian flow conditions by
      \cite{Durlofsky87,KimKarrila,GuazzelliMorris}
\begin{align}
  \label{eq:Rstnc}
    \left(
      \begin{array}{c}
        \FrcHyd \\
        \StsHyd
      \end{array}
    \right)
  =
  - \left(
      \begin{array}{ccc}
        \Rstfv & \Rstfn \\
        \Rstsv & \Rstsn
      \end{array}
    \right)
    \left(
      \begin{array}{c}
        \Vel - \Flv( \Pos ) \\
             - \Stn
      \end{array}
    \right),
\end{align}
  where $\Rstnc$ is the grand resistance matrix.
The elements of this matrix are obtained
  by explicitly solving the flow and pressure fields
    surrounding the moving particle,
  followed by working out their consequences for the particle.
The hydrodynamic force $\FrcHyd$ is
  the zeroth moment of the fluid's deviatoric stress field
      $\StsFl( \PosFl )$ (unit: N/m$^2$)
    integrated over the surface of the particles,
  the deviatoric stress $\StsHyd$ (unit: Nm) is obtained as
    the symmetric first moment of $\StsFl( \PosFl )$.
Analytic solutions of $\Rstnc$ are available
    for spherical and spheroidal particles \cite{KimKarrila};
  the interested reader is referred to the literature
    for details on numerical solvers for colloids of arbitrary shape
      \cite{Makino04a, Garcia07, Aragon11, Palanisamy18}.
% The stress $\StsiHyd{}$ (unit: Nm)
%   is obtained as the symmetric first moments
%     of the deviatoric stress field $\StsFl( \PosFl )$ (unit: N/m$^2$)
%       integrated over the surface of the particles,
%   where the division by volume is omitted
%     to avoid an inconvenient system size dependence in $\Rstnc$.
Under the conditions of Stokesian flow,
% Stokesian flow is governed by force balances:
    the total force on the particle is zero
  and the stress exerted by the particle on the fluid, $\Sts$,
    balances the stress by the fluid on the particle,
% Under the conditions of Stokesian flow,
%   the velocity of a particle experiencing a potential force $\FrcPot$,
%     and the concomitant stress $\Sts$
%       exerted by the particle on the fluid,
%   are obtained from force and stress balances,
\begin{subequations}
\begin{align}
  \label{eq:balanceFrcSmall}
    \FrcPot
  +
    \FrcHyd
 &=
    \Zero,
\\
  \label{eq:balanceSts}
    \Sts
  +
    \StsHyd
 &=
    \Zero.
\end{align}
\end{subequations}
Combining the above equations yields, by partial inversion
  \cite{Durlofsky87,KimKarrila,GuazzelliMorris},
\begin{align}
  \label{eq:MbltyOld}
    \left(
      \begin{array}{c}
        \Veli{} - \Flv \\ % ( \Posi{} ) \\
        \Stsi{}
      \end{array}
    \right)
  =
    \left(
      \begin{array}{ccc}
        \Mblvf & \Mblvn \\
        \Mblsf & \Mblsn
      \end{array}
    \right)
    \left(
      \begin{array}{c}
        \FrcPot \\
        \Stn % \phantom{^\Epot}
      \end{array}
    \right),
\end{align}
  where the grand mobility matrix $\Mblty$
    is related the grand resistance matrix by
\begin{equation}
  \label{eq:PartInv}
    \Mblty
  =
    \left(
      \begin{array}{cc}
        \RstfvI &
        \RstfvI \Rstfn
      \\
        \Rstsv \RstfvI &
        \Rstsv \RstfvI \Rstfn - \Rstsn
      \end{array}
    \right).
\end{equation}
Of the two elements on the {\lhs} of \Eq{eq:MbltyOld},
  the velocity serves as part of an equation of motion, $\Vel = \Posdt$,
    while the stress does not.

%----------------------------------------------------------------------

\subsection{Brownian dynamics}

Consider % next
  an isolated Brownian particle in a quiescent Newtonian fluid.
The evolution of its probability distribution $\pdf( \Pos, t )$
   is described by the Smoluchowski equation
\begin{align}
  \label{eq:SmoluMono}
  \begin{split}
    \dfdx{ \pdf }{ t }
 &=
% - \nabla % \dfdx{ }{ \Posi{} }
%   \cdot
%   \Flux
% =
% - \nabla % dfdx{ }{ \Posi{} }
%   \cdot
%   \Big(
%   - \pdf
%     \Mblty
%     \cdot
%     \nabla % dfdx{ \Epot }{ \Posi{} }
%     \Epot
%   -
%     \Diff
%     \cdot
%     \nabla % dfdx{ \pdf }{ \Posi{} }
%     \pdf
%   \Big)
%\
%&=
% - \nabla % dfdx{ }{ \Posi{} }
%   \cdot
%   \Big(
%   - \pdf
%     \Mblty
%     \cdot
%     \nabla % dfdx{ \Epot }{ \Posi{} }
%     \Epot
%   \Big)
% +
%   \nabla % dfdx{ }{ \Posi{} }
%   \cdot
%   \Big(
%     \Diff
%     \cdot
%     \nabla % dfdx{ \pdf }{ \Posi{} }
%     \pdf
%   \Big)
%\
%&=
% - \nabla % dfdx{ }{ \Posi{} }
%   \cdot
%   \Big(
%   - \pdf
%     \Mblty
%     \cdot
%     \nabla % dfdx{ \Epot }{ \Posi{} }
%     \Epot
%   \Big)
% +
%   \nabla
%   \cdot
%   \Big[
%     \nabla
%     \cdot
%     \Big(
%       \Diff
%       \pdf
%     \Big)
%   -
%     \Big(
%       \nabla
%       \cdot
%       \Diff
%     \Big)
%     \pdf
%   \Big]
%\
%&=
% - \nabla % dfdx{ }{ \Posi{} }
%   \cdot
%   \Big[
%   - \pdf
%     \Mblty
%     \cdot
%     \nabla % dfdx{ \Epot }{ \Posi{} }
%     \Epot
%   +
%     \Big(
%       \nabla
%       \cdot
%       \Diff
%     \Big)
%     \pdf
%   \Big]
% +
%   \nabla
%   \nabla
%   :
%   \Big(
%     \Diff
%     \pdf
%   \Big)
%\
%&=
% - \nabla % dfdx{ }{ \Posi{} }
%   \cdot
%   \Big[
%     \Big(
%     - \Mblty
%       \cdot
%       \nabla % dfdx{ \Epot }{ \Posi{} }
%       \Epot
%     +
%       \nabla
%       \cdot
%       \Diff
%     \Big)
%\
  - \nabla % dfdx{ }{ \Posi{} }
    \cdot
    \Big[
      \Big(
      - \Mblvf
        \nabla % dfdx{ \Epot }{ \Posi{} }
        \Epot
      +
        \nabla
        \cdot
        \Diff
      \Big)
      \pdf
    \Big]
%\
%&\phantom{====}
  +
    \nabla
    \nabla
    :
    \Big(
      \Diff
      \pdf
    \Big),
  \end{split}
\end{align}
  where,
      unlike in \Eq{eq:SmoluGrav},
    it is assumed that the mobility matrix,
        the diffusion matrix $\Diff = \kB T \Mblvf$,
        and the potential
      are functions of the colloidal position.
In this standard form of the Smoluchowksi equation,
  it follows
% one can read off
    from the pre-factor to $\pdf$ in the first term on the {\rhs},
      see Van Kampen \cite{Kampen},
             Eqs~(IX.4.5), (IX.4.11) and (IX.4.12), and
          \"Ottinger \cite{Ottinger}, Eqs~(3.78) and (3.79),
    that the particle experiences,
      in addition to the potential force $\FrcPot = - \nabla \Epot$,
    an \textit{effective} mobility-related force
\begin{align}
  \label{eq:FrcMbl}
    \FrcMbl
  =
    \MblvfI
    \nabla
    \cdot
    \Diff
  =
    \kB T
    \MblvfI
    \nabla
    \cdot
    \Mblvf.
\end{align}
This term arises because
  a first order equation of motion is constructed to describe
    the dynamics resulting from a second order equation of motion
      including a Brownian term with a position-dependent strength.
A tilde is added to emphasize that
  this is not a real force experienced by the particle,
    {\ie} it does not enter a Newtonian equation of motion
      or a second order Langevin equation,
  but an effective force emerging in
    a first order Langevin equation of motion
      on the Smoluchowski time scale.
% Without this term,
%   the dynamics by the first order equation of motion
%     does not recover the Boltzmann distribution.
The equation of motion in the It\^o interpretation then reads as
  \cite{DoiEdwards,Kampen,Ottinger,Briels,Gardiner}
\begin{align}
  \label{eq:EqOfMotion}
    \Posdt
 &=
    \Mblvf
    \left(
      \FrcPot
    +
      \FrcMbl
    +
      \FrcBrwn
    \right),
\end{align}
  where the Brownian force,
      with $\langle \FrcBrwn \rangle = \Zero$,
    obeys the fluctuation-dissipation theorem
\begin{align}
  \label{eq:FlucDiss}
    \left\langle
      \FrcBrwn( t  )
      \dyad
      \FrcBrwn( t' )
    \right\rangle
  =
    2 \kB T
    \Rstfv
    \Dirac( t - t' ).
\end{align}
The It\^o interpretation implies
  that all quantities appearing on the {\rhs} of \Eq{eq:EqOfMotion}
    are evaluated using the positions $\Pos$
      before their incremental change due to the velocity $\Posdt$.
From a physical point of view,
    this Langevin equation only holds true on the Smoluchowski time scale:
  it describes the motion on
    a time scale that
      far exceeds the relaxation time
        of the velocity autocorrelation of the colloid,
          thereby eliminating inertia effects in the force balance,
      but is still short compared to motion over the colloid's size.
The random force entering the dynamics then
  no longer consists of
    an infinite series of uncorrelated (Markovian) delta peaks,
      resulting in discontinuous jumps in the velocity,
  but of a well-defined time-integral over these peaks \cite{Ottinger}.
Note that merely removing the inertial term
    from the Newtonian equation of motion
  is well-known not to yield
    the correct It\^o-form of the first-order Langevin equation of motion
  --~inclusion of the effective force $\FrcMbl$,
      accounting for a bias incurred by Brownian motion
        with a spatially varying mobility matrix,
    is crucial to recovering
      both the correct dynamics and the equilibrium Boltzmann distribution
         \cite{DoiEdwards, Briels}.

The displacements of the particles over a simulation time step $\Dt$
  are usually approximated by the forward Euler scheme
       \cite[see][]{Ottinger},
    {\ie} integrating the {\rhs} of \Eq{eq:EqOfMotion}
      from $t$ to $t + \Dt$ while keeping the coordinates fixed
        at their values at time $t$,
\begin{align}
  \begin{split}
  \label{eq:EqOfMotionDt}
    \Pos( t + \Dt )
  -
    \Pos( t )
%&=
%   \left[
%     \Mblvf
%     \FrcPot
%   +
%     \kB T
%     \nabla
%     \cdot
%     \Mblvf
%   +
%     \Mblvf
%     \FrcBrAv(t)
%   \right]
%   \Dt
%\
 &=
    \Mblvf
    \left[
      \FrcPot
    +
      \FrcBrAv(t)
    \right]
    \Dt
\\
 &\;\;\;
  +
    \kB T
    \nabla
    \cdot
    \Mblvf
    \Dt,
  \end{split}
\end{align}
  where the step-averaged Brownian force
% where the time-averaged (denoted by a bar) Brownian force
      $\FrcBrAv(t) = ( \Dt )^{-1} \int_t^{t+\Dt} \FrcBrwn(\tau) d \tau$
    obeys the fluctuation-dissipation theorem
\begin{align}
  \label{eq:FlucDissDt}
    \left\langle
      \FrcBrAv( t  )
      \dyad
      \FrcBrAv( t' )
    \right\rangle
  =
    2 \kB T
    \Dt
    \Rstfv
    \Kron{t,t'},
\end{align}
  where the Kronecker delta $\Kron{t,t'}$ equals
     one if $t$ and $t'$ refer to the same step and
     zero is $t$ and $t'$ refer to disctinct steps.
In practice,
  the displacement due to the Brownian force is readily calculated as
\begin{align}
% \label{eq:EqOfMotionDt}
  \begin{split}
    \DPosBrwn(t)
  =
    \Mblvf
    \FrcBrAv(t)
    \Dt
  =
    \sqrt{ 2 \kB T }
    \MblvfSq
    \Rndi{}(t)
    \DtSq,
  \end{split}
\end{align}
  where the vector $\Rndi{}(t)$ contains three random numbers
    of zero mean, unit variance and devoid of correlations;
  again, the forces and matrices entering these equations
    are evaluated at time $t$,
      before the position update.
As will be discussed in more detail below,
  there are two Brownian-related terms % $\FrcBrAv$ and $\FrcMbl$
    affecting the displacement in \Eq{eq:EqOfMotionDt}
        and hence the step-averaged velocity of the colloid;
  it then follows from \Eq{eq:Rstnc} that both terms
      contribute to the stress.
Inclusion of a slow (relative to the Smoluchowski time scale) shear flow
  is achieved by
\begin{align}
  \label{eq:EqOfMotionShear}
    \Posdt
% =
%   \frac{
%     \Pos( t + \Dt )
%   -
%     \Pos( t )
%   }{
%     \Dt
%   }
 &=
    \Mblvf
    \left(
      \FrcPot
    +
      \FrcMbl
    +
      \FrcBrwn
    \right)
  +
    \Mblvn
    \Stn
  +
    \Flv,
\end{align}
  and integration from $t$ to $t + \Dt$ extends \Eq{eq:EqOfMotionDt}
    with the flow-related displacement terms
      $- \Mblvn \Stn \Dt + \Flv \Dt$;
  the discretized equation of motion to first order in $\Dt$
    does not contain coupling between diffusion and strain
      \cite{Dotson83,Heyes88}.
The order of strong convergence of this Euler scheme is $1/2$;
  Mil'shtein \cite{Milshtein74, Ottinger} method is required
    to reach an order of one.
The above results are all well-known --
  they form the starting point
    of theoretical developments and Brownian Dynamics simulations
      exploring colloidal dynamics beyond the Smoluchowksi time scale.
Ermak and McCammon \cite{Ermak78}
    derived the above integration scheme % Langevin equation
  starting from the second order Langevin equation of motion
    of a colloid.
The implementation of this scheme
  for a collection of hydrodynamically interacting colloids
%     including their lubrication forces,
    is known as Stokesian Dynamics \cite{Brady88}.
% Brady88 does mention including ln P in the contuum eq of motion,
%   as well as Euler being correct up to dt^2.

%----------------------------------------------------------------------

\subsection{The stress}

The results of the previous two sections are now combined
  to obtain the deviatoric stress on an isolated colloid.
Since we are dealing with a system in Stokesian flow,
  the hydrodynamic force and stress on the colloid
    are obtained by \Eq{eq:Rstnc}.
The hydrodynamic stress is not part of an equation of motion,
  there is no related Fokker-Planck equation,
    nor do the hydrodynamic matrices vary with the stress.
One may therefore apply \Eq{eq:Rstnc},
    still in the It\^o representation
  and keeping in mind that the equation
    is physically valid on the Smoluchowski time scale.
Inserting the velocity derived in \Eq{eq:EqOfMotionShear},
  and using the matrix relations
%   $\Mblvf =   \RstfvI$ and
%   $\Mblvn = - \RstfvI \Rstfn$
    from \Eq{eq:PartInv},
  gives
\begin{align}
  \label{eq:balanceFrcLarge}
    \FrcHyd
%  &=
%   - \Rstfv
%     \left[
%       \PosdtAv
%     -
%       \Flv
%     \right]
%   +
%     \Rstfn
%     \Stn
% \\
%  &=
%   - \Rstfv
%     \left[
%       \Mblvf
%       \left(
%         \FrcPot
%       +
%         \FrcMbl
%       +
%         \FrcBrwn
%       \right)
%     +
%       \Mblvn
%       \Stn
%     \right]
%   +
%     \Rstfn
%     \Stn
% \\
 &=
  - \left(
      \FrcPot
    +
      \FrcMbl
    +
      \FrcBrwn
    \right),
% \\
%      \FrcHydAv
%    +
%      \FrcPot
%    +
%      \FrcMbl
%    +
%      \FrcBrAv
%   &=
%      \Zero,
\end{align}
  thereby recovering the expected force balance
    of a Brownian colloid in Stokesian flow
      on the Smoluchowski time scale.
The effective mobility-related force does not feature
      in a second order Langevin equation of motion of a colloid,
    but emerges in the first order Langevin equation:
  it accounts for
    the average Brownian force on the Smoluchowski time scale
      being non-zero in the presence of a non-constant mobility matrix
        \cite{Ermak78, Kampen, DoiEdwards, Ottinger, Briels}.
% see also \Eq{eq:FrcMbl} and the last term of \Eq{eq:EqOfMotionDt}.
Spatial variations of the mobility matrix affect
    the Brownian displacements of the colloid,
  thereby giving rise to
    an effective force on the Smoluchowski time scale;
  the additional displacement,
    {\ie} the last term in \Eq{eq:EqOfMotionDt},
    contributes to the velocity of the colloid
      and thereby to the stress induced by the colloid.

Returning to \Eq{eq:Rstnc} and
    again inserting the velocity derived in \Eq{eq:EqOfMotionShear},
  one readily obtains
    the deviatoric hydrodynamic stress by the particle on the fluid
      as
\begin{align}
  \label{eq:StressFour}
    \StsHyd
 &=
    \StsPot
  +
    \StsMbl
  +
    \StsBrFr
  +
    \StsStn.
% \\
%  &=
%   - \Rstsv
%     \left[
%       \Mblvf
%       \left(
%         \FrcPot
%       +
%         \FrcMbl
%       +
%         \FrcBrwn
%       \right)
%     +
%       \Mblvn
%       \Stn
%     \right]
%   +
%     \Rstsn
%     \Stn.
\end{align}
Using the matrix relations
%   $\Mblsf = \Rstsv \RstfvI = \Rstsv \Mblvf$ and
%   $\Mblsn = \Rstsv \RstfvI \Rstfn - \Rstsn = \Rstsv \Mblvn - \Rstsn$
    from \Eq{eq:PartInv},
  one finds that the stress consists of
    a potential term
\begin{equation}
    \StsPot
  =
  - \Rstsv
    \Mblvf
    \FrcPot
% =
% - \Rstsv
%   \RstfvI
%   \FrcPot
  =
  - \Mblsf
    \FrcPot
\end{equation}
    and a strain term
\begin{equation}
    \StsStn
  =
  - \left(
      \Rstsv
      \Mblvn
    -
      \Rstsn
    \right)
    \Stn
% =
% - \left(
%   - \Rstsv
%     \RstfvI
%     \Rstfn
%   +
%     \Rstsn
%   \right)
%   \Stn
  =
  - \Mblsn \Stn,
\end{equation}
    both of which already featured in \Eq{eq:MbltyOld},
  as well as two Brownian-related contributions:
    a fluctuating term
\begin{align}
    \StsBrFr
  =
  - \Rstsv
    \Mblvf
    \FrcBrwn
  =
  - \Mblsf
    \FrcBrwn
\end{align}
    and a systematic term
\begin{align}
  \begin{split}
    \StsMbl
 &=
  - \Rstsv
    \Mblvf
    \FrcMbl
  =
  - \Mblsf
    \FrcMbl
\\
 &=
  - \kB T
    \Rstsv
    \nabla
    \cdot
    \Mblvf.
  \end{split}
\end{align}
The latter two terms arise because
  \textit{both the step-averaged Brownian force $\FrcBrwn$,
        with zero mean,
      and the mobility-related effective force $\FrcMbl$,
        accounting for the non-zero mean Brownian displacement
          induced by spatial variations of the mobility matrix,
    contribute to the velocity and displacement of the colloid
      on the Smoluchowski time scale,
        see \Eqs{eq:EqOfMotion} and \eq{eq:EqOfMotionDt},
  while the stress is linear in this velocity
    under Stokesian flow conditions.
}

Combining the force balance of \Eq{eq:balanceFrcLarge}
  with the stress balance of \Eq{eq:balanceSts},
    and repeating the partial inversion of \Eq{eq:MbltyOld},
  one arrives at
\begin{align}
  \label{eq:MbltyNew}
    \left(
      \begin{array}{c}
        \Vel - \Flv \\ % ( \Posi{} ) \\
        \Sts
      \end{array}
    \right)
  =
    \left(
      \begin{array}{ccc}
        \Mblvf & \Mblvn \\
        \Mblsf & \Mblsn
      \end{array}
    \right)
    \left(
      \begin{array}{c}
        \FrcPot + \FrcMbl + \FrcBrwn \\
        \Stn \phantom{^\Epot}
      \end{array}
    \right),
\end{align}
  again in the It\^o interpretation and on the Smoluchowski time scale.
Evaluating this expression recovers
  both the above equation of motion and the four stress contributions.
For simulation purposes,
  forward Euler integration of \Eq{eq:MbltyNew} gives
\begin{subequations}
\begin{align}
% \label{eq:EqOfMotionDt}
    \Pos( t + \Dt )
 &=
    \Pos( t )
  +
    \VelAv( \Pos(t), t )
    \Dt,
\\
    \VelAv( \Pos, t )
 &=
    \Mblvf
    \left[
      \FrcPot
    +
      \FrcMbl
    +
      \FrcBrAv(t)
    \right]
  +
    \Mblvn
    \Stn
  +
    \Flv,
\\
    \StsAv( \Pos, t )
 &=
    \Mblsf
    \left[
      \FrcPot
    +
      \FrcMbl
    +
      \FrcBrAv(t)
    \right]
  +
    \Mblsn
    \Stn,
\end{align}
\end{subequations}
  where the step-averaged velocity $\VelAv$ and stress $\StsAv$
    are determined to the same order in the time step $\Dt$.
This recovers the usual equation of motion,
  with a revised stress.

%----------------------------------------------------------------------

\subsection{Fluctuating contributions to the stress}

The perpetual Brownian motion of colloids affects the stress
  both directly and indirectly.
The direct contributions are represented by the combination of $\StsBrFr$,
  the hydrodynamic stress due to
    colloidal motions induced by the fluctuating Brownian force,
  and $\StsMbl$,
    accounting for a systematic bias in the Brownian force
      whenever the mobility tensor is non-uniform,
        {\eg} due to hydrodynamic interactions between colloids.
The indirect contribution
  results from the combination of Brownian motion
    with potential forces and imposed strain rate,
  collectively determining the time-evolving distribution of the colloids.
Besides these two well-known contributions,
  the multitude of interactions of the colloid
      with the solvent molecules in perpetual thermal motions
    gives rise to two additional stress contributions.
% While the time-averaged zeroth moment and anti-symmetric first moment
%       of the colloid-solvent interaction
%     are known as the usual fluctuating Brownian force and torque,
%       respectively,
%   the time-averaged symmetric first moment
%     causes a fluctuating Brownian stresslet \cite{Palanisamy20}.
The long-time average of the colloid-solvent interaction
  yields the hydrostatic stress $- \pHyd \ONE$,
    with $\pHyd$ the hydrostatic pressure.
Denoting the difference between the short-time and long-time averages
    as the `fluctuating Brownian stresslet,' $\StsBrwn$,
  one finds that $\langle \StsBrwn \rangle = 0$.
Since
  the fluctuating Brownian force and the fluctuating Brownian stresslet
    are distinct projections of the same interactions
      of the colloid with the solvent,
    namely the zeroth moment
      and the symmetric first moment
        of the fluctuating stress field over the colloids surface,
          respectively,
  they are related by a generalized fluctuation-dissipation theorem
    \cite{Palanisamy20},
\begin{align}
  \begin{split}
  \label{eq:FlucDissExt}
  \hspace{-2cm}
 &  \left\langle
      \left(
        \begin{array}{c}
          \FrcBrwn( t  ) \\
          \StsBrwn( t  )
        \end{array}
      \right)
      \dyad
      \left(
        \begin{array}{c}
          \FrcBrwn( t' ) \\
          \StsBrwn( t' )
        \end{array}
      \right)
    \right\rangle
\\
 &\;\;\;\;\;\;\;\;\;
  =
    2 \kB T
    \left(
      \begin{array}{ccc}
        \Rstfv & \Rstfn \\
        \Rstsv & \Rstsn
      \end{array}
    \right)
%   \Rstnc
    \Dirac( t - t' ).
  \end{split}
\end{align}
This coupling does not alter the equation of motion,
  as is readily verified by noting that
    the usual fluctuation-dissipation theorem in \Eq{eq:FlucDiss}
      is a subset of the above expression,
  while the total deviatoric stress $\StsHyd$ in \Eq{eq:StressFour}
    acquires the fluctuating stresslet $\StsBrwn$.

%----------------------------------------------------------------------

\subsection{Angular velocities\label{AngVel}}

Linear flow fields may include a constant rotational component,
    $\Flv(\PosFl) = \FlvO + \Flw \cross \PosFl + \Stn \PosFl$,
  and a colloid in a flow field may acquire an angular velocity $\Av$.
The corresponding extension of \Eq{eq:MbltyNew}
    retains the same concepts,
  while introducing complications
    that we hitherto avoided for clarity of presentation.
For spherical particles,
  it suffices
    to re-interpretate the velocity vectors $\Vel$ and $\Flv$
      as six-vectors combining the linear and angular velocities
        of the colloid and flow field, respectively,
    to re-interpret each of the forces $\FrcPot$, $\FrcMbl$ and $\FrcBrwn$
      as six-vectors combining a force and a torque,
    and
    to extend the grand mobility and grand resistance matrices accordingly.
With these steps,
  the expressions for the motion and stress in \Eq{eq:MbltyNew}
    and the fluctuation-dissipation theorem of \Eq{eq:FlucDissExt}
      hold true again.
% The resultant angular velocity $\Av$ may be discarded for speres.
Only the translational equation of motion needs to be solved
  to explore the evolution of the system in time.
If the particle is non-spherical, however,
  the rotational motion has to be solved as well.
The complication here is
  that the angular velocity $\Av$
    is not the time derivate of a coordinate vector.
One may use Euler angles or a Cartesian rotation vector
  to derive the corresponding mobility matrix and its divergence
    \cite{Evensen08},
  but care must been taken
    to avoid the singular points of the resulting equations of motion.
Furthermore,
  the orientation-dependence of the volume of momentum space
    gives rise to an additional contribution to the torque.
These issues are elegantly solved by using quaternions,
    {\ie} a set of four coordinates
      coupled by a unit-length constraint,
  which results in a remarkably simple equation of motion
%     with an easily solved Lagrange multiplier
      \cite{Ilie15,Palanisamy18}.

%----------------------------------------------------------------------

\subsection{Multiple colloids}

The above derivation is readily extended
    to a collection of $\Ncol$ particles.
Upon re-interpreting $\Posi{}$, $\FrciBrwn{}$, etcetera,
    as vectors comprising all particle coordinates,
      all Brownian forces, etcetera,
  the above equations remain unaltered.
One then obtains for the $i$\supth particle,
\begin{align}
  \begin{split}
  \label{eq:MbltyMany}
    \left(
      \begin{array}{c}
        \Veli{i} \\ % - \Flv( \Posi{i} ) \\
        \Stsi{i}
      \end{array}
    \right)
 &=
    \sum_{ j = 1 }^{ \Ncol }
    \left(
      \begin{array}{ccc}
        \Mblvf^{ij} & \Mblvn^{ij} \\
        \Mblsf^{ij} & \Mblsn^{ij}
      \end{array}
    \right)
    \left(
      \begin{array}{c}
        \FrciPot{j} + \FrciMbl{j} + \FrciBrwn{j} \\
        \Stn \phantom{^\Epot}
      \end{array}
    \right)
\\
 &\;\;\;
  +
    \left(
      \begin{array}{c}
        \Flv( \Posi{i} ) \\
      - \StsiBrwn{i}
      \end{array}
    \right),
  \end{split}
\end{align}
  with fluctuation-dissipation theorem
\begin{align}
  \begin{split}
  \label{eq:FlucDissExtij}
 &  \left\langle
      \left(
        \begin{array}{c}
          \FrciBrwn{i}( t  ) \\
          \StsiBrwn{i}( t  )
        \end{array}
      \right)
      \dyad
      \left(
        \begin{array}{c}
          \FrciBrwn{j}( t' ) \\
          \StsiBrwn{j}( t' )
        \end{array}
      \right)
    \right\rangle
\\
 &\;\;\;\;\;\;\;\;
  =
    2 \kB T
    \left(
      \begin{array}{cc}
        \Rstfv^{ij} & \Rstfn^{ij} \\
        \Rstsv^{ij} & \Rstsn^{ij}
      \end{array}
    \right)
    \Dirac( t - t' ),
  \end{split}
\end{align}
%   Brownian velocities
% \begin{align}
%     \VeliBrwn{}
%   =
%     \sum_{ j = 1 }^{ \Ncol }
%       \Mblvfij
%       \FrciBrwn{j}
% \end{align}
%   Brownian stess
% \begin{align}
%     \StsiBrwn{}
%   =
%     \sum_{ j = 1 }^{ \Ncol }
%       \Mblsfij
%       \FrciBrwn{j}
% \end{align}
%       and a systematic term
% \begin{align}
%     \StsiMbl{}
%   =
%     \Mblsf
%     \FrciMbl{}
%   =
%     \kB T
%     \Rstsv
%     \nabla
%     \cdot
%     \Mblvf.
% \end{align}
  and so on,
    where it should be noted that
      the many-particle matrices $\Mblty$ and $\Rstnc$
          are related by \Eq{eq:PartInv},
        whereas the two-particle matrices $\Mblty^{ij}$ and $\Rstnc^{ij}$
          are not.
In this extension,
  the hydrodynamic matrices account
    for hydrodynamic interactions between the particles,
  {\ie} a force acting on particle $j$ contributes to
    the velocity and stress of particle $i$, and vice versa.
With the potential limited to inter-particle interactions,
    {\ie} in the absence of external interactions,
  the overall deviatoric stress exerted on the fluid,
        $\StsTot$ (unit: N/m$^2$),
      at a strain rate $\Stn$
    is obtained as
\begin{align}
  \label{eq:stressTot}
    \StsTot
  =
    2
    \viscFld
    \Stn
  -
    \frac{ 1 }{ \volBox }
    \sum_i % {i=1}^{\Ncol}
      \Stsi{i}
  -
    \frac{ 1 }{ \volBox }
    \Proj
      \sum_{i<j}
        \Posi{ij}
        \dyad
        \FrciPot{ij},
\end{align}
  where
    the first term on the {\rhs}
      is the stress in the suspending fluid with viscosity $\viscFld$,
    the second term accounts for
      hydrodynamic interactions between the particles and the fluid,
        including fluid-mediated interactions between the particles,
    and the third term is the regular virial expression
      arising from direct inter-particle interactions
          \cite{Batchelor77,DoiEdwards},
        where the projection
\begin{align}
    \Proj \XXX
  =
    ( \XXX + \XXX^\Trnspsd ) / 2 - \det( \XXX ) \ONE
\end{align}
    returns the symmetric traceless part of a matrix $\XXX$.
% This total stress $\StsTot$ is normalized by the volume,
%       recovering the regular unit of N/m$^2$.
The total stress tensor of the suspension is obtained
  by adding the hydrostatic pressure $- \pHyd \ONE$ to \Eq{eq:stressTot}.
The contribution of the colloids to this total stress,
    also known as the osmotic stress,
  is obtained from \Eq{eq:stressTot} by
    removing
      the bulk term $2 \viscFld \Stn$ and
      the projection $\Proj$,
    and adding
      the kinetic contribution $- \Ncol \kB T \ONE / \volBox$.

%======================================================================

\section{Comparison of stress expressions\label{sec:comparison}}

Comparing the novel expression for the stress %  with Batchelor's,
  with earlier expressions, by
  Batchelor \cite{Batchelor77}, Eqs.~(2.2) and (3.10),
  Felderhof \cite{Felderhof87b}, Eq.~(7.17),
  Brady \cite{Brady93}, Eqs~(38) through (40), and
  N\"agele and Bergenholtz \cite{Nagele98}, Eq.~(31),
  reveals a number of similarities and differences.
% The motion of the particles induced by potential forces
%     is seen to contribute to the stress,
%   whereas this was not the case in the conventional expressions.
% Our expression reproduces
%   the usual fluid, convective, potential and kinetic contributions
%     to the stress,
%   as previously derived
%     by \citet{Felderhof87b}[Eq.~(7.17)]
%     and \citet{Brady93}[Eqs~(38) through (40)].
The contributions due to the strain are identical,
  where the stress $\StsiStn{i}$ in \Eq{eq:Batchelor2}
    is understood to include hydrodynamic interactions
      between the colloids,
  $\StsiStn{i} = - \sum_j \Mblsn^{ij} \Stn$.
From the velocity relation under pure strain \cite{Batchelor77,Nagele98},
\begin{align}
    \Veli{i}
  =
    \Flv( \Posi{i} )
  +
    \Stn
    :
    \CCi
  =
    \Flv( \Posi{i} )
  +
    \sum_j
      \Mblvn^{ij}
      :
      \Stn,
\end{align}
  follows $\CCi = \sum_j \Mblsf^{ji}$,
    where a symmetry rule of the grand mobility matrix
        \cite{Makino04a}
      was used in the last step.
The potential-induced hydrodynamic term in \Eq{eq:Nagele} then reads as
  $\StsPot = - \sum_i \CCi \cdot \FrciPot{i}
           = - \sum_{ij} \Mblsf^{ji} \FrciPot{i}$,
  in agreement with the corresponding term
%     $- \sum_{ij} \Mblsf^{ij} \FrciPot{j}$,
    in \Eq{eq:MbltyMany}.

The differences % between the current and previous stress expressions
  are in the stress contributions by the Brownian forces.
Previous derivations of stress expressions
  are based on the assumption that
    the mean contribution of the Brownian force
      is provided by the thermodynamic force $\FrcThrm$,
    giving rise to a thermodynamic stress term $\StsThrm$.
% \begin{align}
%     \StsThrm
%   =
%     \frac{ \kB T }{ \volBox }
%     \sum_i % {i=1}^{\Ncol}
%       \left(
%         \CCi
%         +
%         \Posi{i}
%         \ONE
%       \right)
%       \cdot
%       \ln \pdf,
% \end{align}
%   while this effective force is well-known
%     not to enter the equation of motion of the colloids.
Here, instead,
  the mean contribution of the Brownian force is equated to
     the mobility-related effective force $\FrcMbl$, see \Eq{eq:FrcMbl}.
%  which also features in the first-order Langevin equation of motion
%    and is routinely applied
%      in Brownian and Stokesian Dynamics simulations.
These two effective forces are fundamentally different,
    being based on the probability distribution
      and the grand mobility matrix, respectively,
  and consequently the corresponding stresses have little in common.
% It is therefore not possible
%   to derive a relation between the two corresponding stress terms.
After rewriting the thermodynamic stress term
 to eliminate the probability distribution,
    as in \Eqs{eq:Batchelor2} and \eq{eq:Nagele},
  both stress terms acquire superficial similarities
    as divergences of segments of the hydrodynamic matrices,
\begin{align}
    \StsThrm
  =
  - \frac{ \kB T }{ \volBox }
    \sum_i
      \nabla_i
      \cdot
      \CCi
  =
  - \frac{ \kB T }{ \volBox }
    \sum_{ij}
      \nabla_i
      \cdot
      \Mblsf^{ij}
\end{align}
and
\begin{align}
    \StsMbl
  =
  - \frac{ 1 }{ \volBox }
    \sum_{ij}
      \Mblsf^{ij}
      \FrciMbl{j}
  =
  - \frac{ \kB T }{ \volBox }
    \sum_{ijk}
      \Rstsv^{ik}
      \nabla_j
      \cdot
      \Mblvf^{kj},
\end{align}
  respectively.
Since $\Mblsf = \Rstsv \Mblvf$,
    as follows from \Eq{eq:PartInv},
  the two stresses are different in general.
A rare exception is a dispersion consisting of a single sphere,
  in which case $\Mblsf = \Rstsv = \ZERO$.

A second difference
  is the explicit inclusion of all fluctuating Brownian contributions
    in \Eq{eq:MbltyMany}.
Their presence allows for a self-consistency test by comparing
  the viscosity obtained
    from the average stress at constant low shear rate
  with the viscosity extracted
    from the thermal stress fluctuations in equilibrium
      using the Green-Kubo formalism.
It is not possible
  to conclude that a stress expression passes this test
    based on an analysis
      that bypasses the fluctuating Brownian stress contributions
        and their correlations to the colloidal dynamics
        \cite{Nagele98,Palanisamy20}.
The consistency test of the revised stress expression
  is a topic of ongoing research.
Note that identical forces enter
    both the equation of motion and the novel stress expression,
  because the non-straining part of the hydrodynamic stress
    is a consequence of
      the motion of the colloids relative to the flow field,
        see \Eq{eq:Rstnc}.
In applications of Batchelor's approach,
    besides the ommitted fluctuating Brownian contributions to the stress,
  the effective mobility-related force $\FrcMbl$
    is used in the equation of motion
      but not in the stress,
  while the effective thermodynamic force $\FrcThrm$
    features in the stress,
\begin{align}
%   \left\langle
      \StsThrm
%   \right\rangle
  =
%   \left\langle
    - \frac{ 1 }{ \volBox }
      \sum_{ij}
        \Mblsf^{ij}
        \FrciThrm{j},
%   \right\rangle
% =
%   \left\langle
%   - \frac{ \kB T }{ \volBox }
%     \sum_{ij}
%       \nabla_i
%       \cdot
%       \Mblsf^{ij},
%   \right\rangle,
\end{align}
  but does not appear in the equation of motion.

%======================================================================

\section{Discussion and conclusions\label{sec:DiscConc}}

The `thermodynamic force' $\FrcThrm$
  was presented by Batchelor \cite{Batchelor76} as
    `an alternative and much simpler method
        for the statistical mechanics part of the investigation
       which is a generalization of the argument used by Einstein
         and which gives the asymptotic or long-time statistical properties
           of the displacement of particles
             in terms of the thermal energy of the medium.'
Batchelor \cite{Batchelor77}, and many authors since,
  have used the thermodynamic force as
    the average resultant of the Brownian force,
        driving the colloids against the concentration gradient,
      in the derivation of stress expressions for colloidal suspensions.
But the average resultant of the Brownian force is well-known
 to derive from the divergence of the mobility matrix,
   referred to above as the mobility-related force $\FrcMbl$.
A new stress expression was derived,
%   see \Eqs{eq:StressFour}, \eq{eq:MbltyMany},
%        \eq{eq:FlucDissExtij} and \eq{eq:stressTot},
    see \Eq{eq:StressFour}
      and \Eqs{eq:MbltyMany} through \eq{eq:stressTot},
  in which the Brownian forces entering the stress calculation
    match those entering the equation of motion of the colloids.

The interpretation of the thermodynamic force as
  the average contribution of Brownian motion
    is widespread in the literature on the stress of colloidal suspensions.
Bossis and Brady \cite{Bossis89} are a rare exception,
  by presenting an alternative derivation of a stress expression
    without making use of the thermodynamic force
      and including the mobility-related force instead.
% \citet{Bossis89} derived the same $\StsiThrm{i}$
They replace \Eq{eq:EqOfMotion} by
  an alternative first order Langevin equation of motion
      that by an averaged Mil'shtein approximation
        recovers \Eq{eq:EqOfMotionDt} upon integration;
  the stress then follows by integration
    of $\Rstsv \Posdt$
      over the time step,
        again by averaging the Mil'shtein approximation
          over the Brownian forces.
This derivation is hampered, however,
  by building on the assumption
    that \Eq{eq:EqOfMotionDt} is the correct expression
      and \Eq{eq:EqOfMotion} an approximation~--
  it is well-known that \Eq{eq:EqOfMotion} is the correct expression
      in the It\^o interpretation
    while \Eq{eq:EqOfMotionDt}
      is an approximation to its integration over a time step
        \cite{Ermak78,Kampen,Ottinger}.

More work is needed to establish the impact
    of the revised averaged Brownian contribution
  on the stress and viscosity calculations of the past 40 years.
The good agreement between simulation results and experimental data
  on the viscosity of suspensions of spherical particles,
    see for instance the variation of viscosity with volume fraction
      reported by Foss and Brady \cite{Foss00},
  indicates that the mean Brownian term
    makes a relatively minor contribution in this particular case.
A number of simulations and derivations should be repeated carefully
  to establish the particular consequences for other systems.
%   {\eg} polymers \cite{DoiEdwards}[Eq.~(4.130)].
Numerical results illustrating the impact on simple colloidal systems
  will be presented in a forthcoming publication.

%======================================================================

\acknowledgments

This work is part of
  the Computational Sciences for Energy Research
    Industrial Partnership Programme
  co-financed by Shell Global Solutions B.V.
  and the Netherlands Organisation for Scientific Research (NWO).

%======================================================================

% \appendix

% \section{Appendix: Deformable particles}\label{app:deform}

%----------------------------------------------------------------------

% \bibliography{manus}

\end{document}